\input harvmac
\input epsf

\def\scri{{\cal{I}}}
\def\bx{{\vec{x}}}
\def\p{\partial}
\def\msurr{\mathsurround=0pt} 
\def\overleftrightarrow#1{\vbox{\msurr\ialign{##\crcr
        $\leftrightarrow$\crcr\noalign{\kern-1pt\nointerlineskip}
        $\hfil\displaystyle{#1}\hfil$\crcr}}}                    
\def\vy{{\vec{y}}}
\def\vq{{\vec{q}}}
\def\dst{dS${}_3$}
\def\vp{{\vec{p}}}
\def\vx{{\vec{x}}}

\newcount\figno
\figno=0 
\def\fig#1#2#3{
\par\begingroup\parindent=0pt\leftskip=1cm\rightskip=1cm\parindent=0pt
\baselineskip=11pt
\global\advance\figno by 1
\midinsert
\epsfxsize=#3
\centerline{\epsfbox{#2}}
\vskip 12pt
{\bf Fig.\ \the\figno: } #1\par
\endinsert\endgroup\par
}
\def\figlabel#1{\xdef#1{\the\figno}}

\lref\dscftgeneral{
F.~L.~Lin and Y.~S.~Wu,
``Near-horizon Virasoro symmetry and the entropy of de Sitter space in
any dimension,''
Phys.\ Lett.\ B {\bf 453}, 222 (1999)
[arXiv:hep-th/9901147];
D.~Klemm,
``Some aspects of the de Sitter/CFT correspondence,''
arXiv:hep-th/ 0106247;
J.~Bros, H.~Epstein and U.~Moschella,
``The asymptotic symmetry of de Sitter spacetime,''
arXiv:hep-th/0107091;
S.~Cacciatori and D.~Klemm,
``The asymptotic dynamics of de Sitter gravity in three dimensions,''
arXiv:hep-th/0110031;
B.~McInnes,
``Exploring the similarities of the dS/CFT and AdS/CFT correspondences,''
arXiv:hep-th/0110062;
V.~Balasubramanian, J.~de Boer and D.~Minic,
``Mass, entropy and holography in asymptotically de Sitter spaces,''
arXiv:hep-th/0110108;
B.~G.~Carneiro da Cunha,
``Three-dimensional de Sitter gravity and the correspondence,''
arXiv:hep-th/0110169;
R.~G.~Cai,
``Cardy-Verlinde formula and asymptotically de Sitter spaces,''
arXiv:hep-th/0111093;
M.~H.~Dehghani,
``Kerr-de Sitter spacetimes in various dimension and dS/CFT
correspondence,''
arXiv:hep-th/0112002.
B.~McInnes,
``The dS/CFT correspondence and the big smash,''
arXiv:hep-th/0112066.
Y.~S.~Myung,
``Dynamic dS/CFT correspondence using the brane cosmology,''
arXiv:hep-th/0112140.
S.~Nojiri and S.~D.~Odintsov,
``Wilson loop and dS/CFT correspondence,''
arXiv:hep-th/0112152.
}

\lref\BanksYP{
T.~Banks and W.~Fischler,
``M-theory observables for cosmological space-times,''
arXiv:hep-th/0102077.
}

\lref\SachsQB{
I.~Sachs and S.~N.~Solodukhin,
``Horizon holography,''
Phys.\ Rev.\ D {\bf 64}, 124023 (2001)
[arXiv:hep-th/0107173].
}

\lref\desittergeneral{
J.~Maldacena and A.~Strominger,
``Statistical entropy of de Sitter space,''
JHEP {\bf 9802}, 014 (1998)
[arXiv:gr-qc/9801096].
R.~Bousso,
``Proliferation of de Sitter space,''
Phys.\ Rev.\ D {\bf 58}, 083511 (1998)
[arXiv:hep-th/9805081].
M.~I.~Park,
``Statistical entropy of three-dimensional Kerr-de Sitter space,''
Phys.\ Lett.\ B {\bf 440}, 275 (1998)
[arXiv:hep-th/9806119].
C.~M.~Hull,
``Timelike T-duality, de Sitter space, large N gauge theories and  topological field theory,''
JHEP {\bf 9807}, 021 (1998)
[arXiv:hep-th/9806146].
M.~Banados, T.~Brotz and M.~E.~Ortiz,
``Quantum three-dimensional de Sitter space,''
Phys.\ Rev.\ D {\bf 59}, 046002 (1999)
[arXiv:hep-th/9807216].
W.~T.~Kim,
``Entropy of 2+1 dimensional de Sitter space in terms of brick wall  method,''
Phys.\ Rev.\ D {\bf 59}, 047503 (1999)
[arXiv:hep-th/9810169].
R.~Bousso,
``Quantum global structure of de Sitter space,''
Phys.\ Rev.\ D {\bf 60}, 063503 (1999)
[arXiv:hep-th/9902183].
S.~Hawking, J.~Maldacena and A.~Strominger,
``DeSitter entropy, quantum entanglement and AdS/CFT,''
JHEP {\bf 0105}, 001 (2001)
[arXiv:hep-th/0002145].
R.~Bousso,
``Bekenstein bounds in de Sitter and flat space,''
JHEP {\bf 0104}, 035 (2001)
[arXiv:hep-th/0012052].
M.~K.~Parikh and S.~N.~Solodukhin,
``De Sitter brane gravity: From close-up to panorama,''
Phys.\ Lett.\ B {\bf 503}, 384 (2001)
[arXiv:hep-th/0012231].
A.~Volovich,
``Discreteness in deSitter space and quantization of Kaehler manifolds,''
arXiv:hep-th/0101176.
A.~Chamblin and N.~D.~Lambert,
``de Sitter space from M-theory,''
Phys.\ Lett.\ B {\bf 508}, 369 (2001)
[arXiv:hep-th/0102159].
V.~Balasubramanian, P.~Horava and D.~Minic,
``Deconstructing de Sitter,''
JHEP {\bf 0105}, 043 (2001)
[arXiv:hep-th/0103171].
M.~Li,
``Matrix model for de Sitter,''
arXiv:hep-th/0106184.
Y.~h.~Gao,
``Symmetries, matrices, and de Sitter gravity,''
arXiv:hep-th/0107067.
S.~Nojiri and S.~D.~Odintsov,
``Quantum cosmology, inflationary brane-world creation 
and dS/CFT  correspondence,''
arXiv:hep-th/0107134.
E.~Halyo,
``De Sitter entropy and strings,''
arXiv:hep-th/0107169.
J.~W.~Moffat,
``M-theory and de Sitter space,''
arXiv:hep-th/0107183.
R.~Kallosh,
``N = 2 supersymmetry and de Sitter space,''
arXiv:hep-th/0109168.
C.~M.~Hull,
``de Sitter space in supergravity and M theory,''
JHEP {\bf 0111}, 012 (2001)
[arXiv:hep-th/0109213].
C.~M.~Hull,
``Domain wall and de Sitter solutions of gauged supergravity,''
arXiv:hep-th/0110048.
S.~Nojiri and S.~D.~Odintsov,
``De Sitter space versus Nariai black hole: Stability in d5 higher  derivative gravity,''
arXiv:hep-th/0110064.
R.~Kallosh, A.~D.~Linde, S.~Prokushkin and M.~Shmakova,
``Gauged supergravities, de Sitter space and cosmology,''
arXiv:hep-th/0110089.
T.~Shiromizu,
``Gravitational mass in asymptotically de Sitter space-times with
compactified dimensions,''.
R.~G.~Cai, Y.~S.~Myung and Y.~Z.~Zhang,
``Check of the mass bound conjecture in de Sitter space,''
arXiv:hep-th/0110234.
U.~H.~Danielsson,
``A black hole hologram in de Sitter space,''
arXiv:hep-th/0110265.
S.~Ogushi,
``Holographic entropy on the brane in de Sitter Schwarzschild space,''
arXiv:hep-th/0111008.
A.~M.~Ghezelbash and R.~B.~Mann,
``Action, mass and entropy of Schwarzschild-de Sitter black holes  and the de Sitter / CFT correspondence,''
arXiv:hep-th/0111217.
G.~W.~Gibbons and C.~M.~Hull,
``de Sitter space from warped supergravity solutions,''
arXiv:hep-th/0111072.
D.~Youm,
``The Cardy-Verlinde formula and asymptotically de Sitter brane universe,''
arXiv:hep-th/0111276.
M.~Cvetic, S.~Nojiri and S.~D.~Odintsov,
``Black Hole Thermodynamics and Negative Entropy in deSitter and Anti-deSitter Einstein-Gauss-Bonnet gravity,''
arXiv:hep-th/0112045.
P.~Berglund, T.~Hubsch and D.~Minic,
``de Sitter spacetimes from warped compactifications of IIB string  theory,''
arXiv:hep-th/0112079.
}

\lref\Bjorken{
J.~D.~Bjorken and S.~D.~Drell,
{\it Relativistic Quantum Fields}, McGraw-Hill (1965).}

\lref\PolchinskiRY{
J.~Polchinski,
``S-matrices from AdS spacetime,''
arXiv:hep-th/9901076.
}

\lref\ArefevaMI{
I.~Y.~Aref'eva,
``On the holographic S-matrix,''
arXiv:hep-th/9902106.
}

\lref\SusskindVK{
L.~Susskind,
``Holography in the flat space limit,''
arXiv:hep-th/9901079.
}

\lref\fulling{
S.~A.~Fulling, {\it Aspects of Quantum Field Theory in
Curved Space-Time}, Cambridge University Press (1989).}

\lref\GiddingsQU{
S.~B.~Giddings,
``The boundary S-matrix and the AdS to CFT dictionary,''
Phys.\ Rev.\ Lett.\  {\bf 83}, 2707 (1999)
[arXiv:hep-th/9903048].
}

\lref\WittenQJ{
E.~Witten,
``Anti-de Sitter space and holography,''
Adv.\ Theor.\ Math.\ Phys.\  {\bf 2}, 253 (1998)
[arXiv:hep-th/9802150].
}

\lref\GubserBC{
S.~S.~Gubser, I.~R.~Klebanov and A.~M.~Polyakov,
``Gauge theory correlators from non-critical string theory,''
Phys.\ Lett.\ B {\bf 428}, 105 (1998)
[arXiv:hep-th/9802109].
}

\lref\MaldacenaRE{
J.~Maldacena,
``The large $N$ limit of superconformal field theories and supergravity,''
Adv.\ Theor.\ Math.\ Phys.\  {\bf 2}, 231 (1998)
[Int.\ J.\ Theor.\ Phys.\  {\bf 38}, 1113 (1998)]
[arXiv:hep-th/9711200].
}

\lref\ChernikovZM{
N.~A.~Chernikov and E.~A.~Tagirov,
``Quantum Theory Of Scalar Fields In De Sitter Space-Time,''
Annales Poincare Phys.\ Theor.\ A {\bf 9}, 109 (1968).
}

\lref\TagirovVV{
E.~A.~Tagirov,
``Consequences Of Field Quantization In De Sitter Type Cosmological Models,''
Annals Phys.\  {\bf 76}, 561 (1973).
}

\lref\BalasubramanianSN{
V.~Balasubramanian, P.~Kraus and A.~E.~Lawrence,
``Bulk vs. boundary dynamics in anti-de Sitter spacetime,''
Phys.\ Rev.\ D {\bf 59}, 046003 (1999)
[arXiv:hep-th/9805171].
}

\lref\BreitenlohnerJF{
P.~Breitenlohner and D.~Z.~Freedman,
``Stability In Gauged Extended Supergravity,''
Annals Phys.\  {\bf 144}, 249 (1982).
}

\lref\AllenUX{
B.~Allen,
``Vacuum States In De Sitter Space,''
Phys.\ Rev.\ D {\bf 32}, 3136 (1985).
}

\lref\MottolaAR{
E.~Mottola,
``Particle Creation In De Sitter Space,''
Phys.\ Rev.\ D {\bf 31}, 754 (1985).
}

\lref\TolleyGG{
A.~J.~Tolley and N.~Turok,
``Quantization of the massless minimally coupled scalar field and the
dS/CFT correspondence,''
arXiv:hep-th/0108119.
}

\lref\StromingerGP{
A.~Strominger,
``Inflation and the dS/CFT correspondence,''
JHEP {\bf 0111}, 049 (2001)
[arXiv:hep-th/0110087].
}

\lref\SpradlinPW{
M.~Spradlin, A.~Strominger and A.~Volovich,
``Les Houches lectures on de Sitter space,''
arXiv:hep-th/0110007.
}

\lref\StromingerPN{
A.~Strominger,
``The ds/CFT correspondence,''
JHEP {\bf 0110}, 034 (2001)
[arXiv:hep-th/0106113].
}

\lref\BalasubramanianRI{
V.~Balasubramanian, S.~B.~Giddings and A.~E.~Lawrence,
``What do CFTs tell us about anti-de Sitter spacetimes?,''
JHEP {\bf 9903}, 001 (1999)
[arXiv:hep-th/9902052].
}

\lref\BalasubramanianDE{
V.~Balasubramanian, P.~Kraus, A.~E.~Lawrence and S.~P.~Trivedi,
``Holographic probes of anti-de Sitter space-times,''
Phys.\ Rev.\ D {\bf 59}, 104021 (1999)
[arXiv:hep-th/9808017].
}

\lref\MincesZY{
P.~Minces and V.~O.~Rivelles,
``Energy and the AdS/CFT correspondence,''
arXiv:hep-th/0110189.
}

\lref\KlebanovTB{
I.~R.~Klebanov and E.~Witten,
``AdS/CFT correspondence and symmetry breaking,''
Nucl.\ Phys.\ B {\bf 556}, 89 (1999)
[arXiv:hep-th/9905104].
}

\lref\HellermanYI{
S.~Hellerman, N.~Kaloper and L.~Susskind,
``String theory and quintessence,''
JHEP {\bf 0106}, 003 (2001)
[arXiv:hep-th/0104180].
}

\lref\WittenKN{
E.~Witten,
``Quantum gravity in de Sitter space,''
arXiv:hep-th/0106109.
}

\lref\FischlerYJ{
W.~Fischler, A.~Kashani-Poor, R.~McNees and S.~Paban,
``The acceleration of the universe, a challenge for string theory,''
JHEP {\bf 0107}, 003 (2001)
[arXiv:hep-th/0104181].
}

\lref\BMS{
R.~Bousso, A.~Maloney and A.~Strominger,
``Conformal Vacua and Entropy in de Sitter Space,''
to appear.}

\lref\gr{
I.~S.~Gradshteyn and I.~M.~Ryzhik,
{\it Table of Integrals, Series, and Products},
Academic Press (2000).}

\lref\fred{
D.~Z.~Freedman, S.~D.~Mathur, A.~Matusis and L.~Rastelli,
``Correlation functions in the CFT($d$)/AdS($d+1$) correspondence,''
Nucl.\ Phys.\ B {\bf 546}, 96 (1999)
[arXiv:hep-th/9804058].
}

\Title{\vbox{\baselineskip12pt
        \hbox{hep-th/0112223}
        \hbox{PUTP-2017}
}}{Vacuum States and the $S$-Matrix in dS/CFT}

\centerline{
Marcus Spradlin${}^{1}$ and Anastasia Volovich${}^{2}$
}

\bigskip
\centerline{${}^{1}$~Department of Physics}
\centerline{Princeton University}
\centerline{Princeton, NJ 08544}
\centerline{\tt spradlin@feynman.princeton.edu}
\centerline{}
\centerline{${}^{2}$~Department of Physics}
\centerline{Harvard University}
\centerline{Cambridge, MA 02138}
\centerline{\tt nastya@gauss.harvard.edu}

\vskip .3in
\centerline{\bf Abstract}
We propose a definition of dS/CFT correlation functions
by equating them to $S$-matrix elements for  scattering 
particles from $\scri^-$ to $\scri^+$.  
In planar coordinates, which cover half of de Sitter space,
we consider instead the $S$-vector obtained by specifying
a fixed state on the horizon.
We construct the one-parameter family of de Sitter invariant
vacuum states for a massive scalar field in these
coordinates, and show that the vacuum obtained by analytic continuation
from the sphere has no particles on the past horizon.
We use this formalism to
provide evidence that the  one-parameter family of vacua corresponds to
marginal deformations of the CFT by computing
a three-point function.

\smallskip

\Date{}

\listtoc
\writetoc

\newsec{Introduction}

Understanding quantum gravity in
de Sitter space remains one of the most important problems in
theoretical physics.
A correspondence relating gravity in de Sitter space
to a conformal field theory has recently been suggested \StromingerPN\ (see
also \WittenKN), and subsequently studied by several authors
\dscftgeneral.
Recent works on de Sitter space include \desittergeneral.

The dS/CFT correspondence is modeled in analogy with the
AdS/CFT correspondence \refs{\MaldacenaRE, \GubserBC, \WittenQJ},
which has proven to be phenomenally successful.
But it is important to keep in mind
that in the prehistoric days of AdS/CFT, when the first signs were
emerging 
that there might be some connection between supergravity on AdS space
and conformal field theories, it would have seemed beyond hope
to expect that these developments
would lead to a nonperturbative definition of
quantum gravity on AdS space, and to all of the remarkable advances
that have been made in our understanding of gauge theories.
AdS/CFT turned out to be more wonderful than we had any right to
expect, so we should not be prejudiced against dS/CFT 
simply because it has some mysterious and confusing aspects and has
not yet borne the rich fruit of its AdS brother.

Therefore we proceed modestly in this paper, by elucidating the
connection between gravity on de Sitter space and  conformal field theory
correlation functions.
Our probe will be an interacting
real scalar field of mass $m$.
This turns out to be more interesting than it might seem at first since
it is known that there is no unique de Sitter invariant vacuum state
for a massive scalar field, but instead a family of vacua labelled
by a complex parameter $\gamma$.
Changing the vacuum $|\gamma \rangle$
in the bulk of \dst\ has been argued to correspond to a marginal
deformation of the associated CFT \BMS.

The central result of this paper is a proposal for how to extract
CFT correlation functions from $n$-point correlation functions of the scalar
field on \dst.  Along the way we highlight
the important differences between dS and AdS which make naive
extrapolation of some AdS/CFT results problematic.
In the global picture of de Sitter, there are four
CFT operators associated to the scalar field $\phi$, which are labelled
${\cal{O}}_\pm^{\rm in,out}$, and have weights $h_\pm = 1 \pm \sqrt{1-m^2}$.
Only two of these operators are independent, and in general the out
operators can be related to the in operators by path integral evolution
from $\scri^-$ to $\scri^+$. 
We equate
correlation functions of ${\cal{O}}^{\rm in,out}_+$ with
$S$-matrix elements for particles coming in from $\scri^-$ and going
out to $\scri^+$\foot{Our $S$-matrix is the standard
one of perturbative quantum field theory, as distinct from
the (finite-dimensional) matrices of \refs{\WittenKN, \BanksYP}, although
it would be very interesting to understand a connection with these works.}.
This definition of CFT correlation functions
is motivated by a similar construction for AdS which has been
developed in \refs{\PolchinskiRY,
\SusskindVK,\BalasubramanianRI,
\ArefevaMI, 
\GiddingsQU}.
We do not address the important issue that
these $S$-matrix elements are only `meta-observables' and cannot be
probed by any single observer in de Sitter space
\refs{\HellermanYI, \FischlerYJ, \WittenKN}.

In planar coordinates, which only cover half of de Sitter space, one
has only half as many operators.  For example, in the patch ${\cal{O}}^-$
which includes the causal past of an observer sitting at the south pole,
there are no asymptotic out states, so the best one can do is
to study the $S$-vector \refs{\WittenKN, \HellermanYI}.
This leads to a natural definition of correlation functions
involving
only
${\cal{O}}^{\rm in}_\pm$.
Along the way, we prove
the somewhat surprising result that the Euclidean vacuum state (which is
the one obtained by analytic continuation from the sphere to de Sitter
spacetime) is the state with no particles on the
horizon.

The plan of the paper is the following.
In section 2 we introduce global and planar coordinate systems for \dst, 
mode expansions for the scalar field, and the bulk-boundary propagators.
In section 3 we review the construction of the de Sitter invariant
vacuum states $|\gamma\rangle$ in global coordinates and  
record the  two-point functions of the scalar field.
In section 4 we
show how these vacuum states can be obtained naturally in
planar coordinates as well, and that the Euclidean vacuum
is the one with no particles on the horizon.
Section 5 contains the general prescription for calculating
CFT correlation functions in global coordinates and explains
the connection to the $S$-matrix.
At the end of section 5 we outline the calculation of a CFT three-point
function and show that the parameter $\gamma$ appears nontrivially
in an invariant ratio of correlation functions,
providing an evidence that these vacua are marginal
deformations of the associated CFT.
The prescription for dS/CFT correlation functions
in planar coordinates appears in section 6, where the motivation
is provided by the $S$-vector.

\newsec{Coordinates, Modes and Bulk-Boundary Propagators}

In this paper we consider
an interacting scalar field $\phi$
in \dst\ with the action
\eqn\action{
S = - {1 \over 2} \int \sqrt{-g} \ \left[(\nabla \phi)^2 + m^2 \phi^2+
V(\phi)\right].
}
We  set the de Sitter radius $l$ to unity and assume that $m^2 > 1$.  This
condition, while not essential,
simplifies the discussion for reasons that will become
clear shortly.
Most of the results of this paper generalize more or less
straightforwardly to scalars with $m^2\le 1$, higher spin fields, and
higher dimensional de Sitter space.
We will comment on
exceptions to this expectation as they arise.

We consider 
two coordinate systems:  global coordinates $(\tau,\Omega)$
and planar coordinates $(t,\vx)$.
Here $\Omega$ is a point on $S^2$ and $\vx$ is a point on ${\bf R}^2$.
The metric is
\eqn\metric{
ds^2 = - d\tau^2 + \cosh^2\tau\, d\Omega_2^2
= {1 \over t^2}(-dt^2 + d\vx^2).
}
Global coordinates cover all of \dst, with $\tau$ running from $-\infty$
on $\scri^-$ to $+\infty$ on $\scri^+$, while planar coordinates only
cover the causal past of an observer on the south pole.  In planar
coordinates $\scri^-$ is at $t=0$ and the horizon lies at $t=+\infty$.
A number of additional coordinate systems
and further  details can be found in \SpradlinPW.

\fig{The Penrose diagram for de Sitter space.  $(a)$ Global coordinates
cover all of de Sitter space, with dotted lines signifying slices of
constant $\tau$, which ranges from $-\infty$ to $+\infty$.  $(b)$
Planar coordinates
cover only the causal past ${\cal{O}}^-$ of an observer at the
south pole.  The dotted lines are lines of constant $t$, with $t = 0$
on $\scri^-$ and $t \to +\infty$ at the horizon.}
{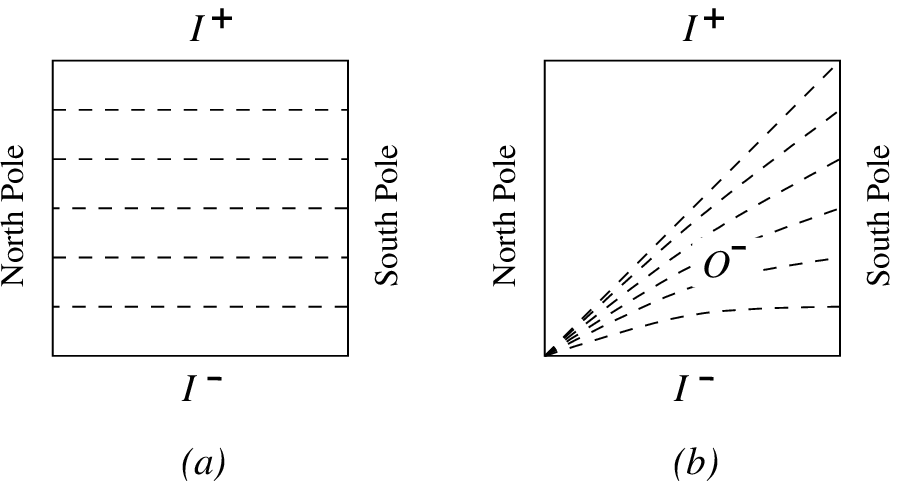}{3.7in}

For some purposes, particularly cosmology, the
region ${\cal{O}}^+$ corresponding to the causal future
of an observer at the south pole may be of greater interest than
${\cal{O}}^-$ (see \StromingerGP\ for an interesting application).
All of the formulas presented in this paper can be adapted to
${\cal{O}}^+$ by taking $t \to -t$, so that $t=0$ corresponds to
$\scri^+$ and $t=-\infty$ corresponds to the horizon.

In global coordinates we will make use of the antipodal map on \dst.
Actually there are two antipodal maps, one which just takes $\Omega$
to the antipodal point on $S^2$, and one which in addition takes
$\tau \to -\tau$.  We will use the notation $\Omega_{\rm A}$ for
the former and $x_{\rm A}$ for the latter, where $x = (\tau,\Omega)$.

The following two subsections catalog the mode expansions for a
free scalar field and introduce the bulk-boundary propagators in
the two coordinate systems.

\subsec{Global Coordinates}

In global coordinates we follow closely the conventions of \BMS.
A basis of positive frequency solutions of the free Klein-Gordon equation
is given by
\eqn\globalmodes{
\phi_{lm}(\tau,\Omega) = y_l(\tau) Y_{lm}(\Omega),
}
where
\eqn\ydef{
y_l(\tau) = e^{i \theta_l}
\sqrt{2 \over \mu} (1 + e^{2 \tau})^l e^{(1 - i \mu)\tau}
F(l{+}1, l{+}1{-}i\mu, 1{-}i \mu,-e^{2 \tau}),
}
and the phase\foot{The reader may well wonder
why we have bothered to introduce such a complicated phase,
since  the overall phase of a mode function is of course
irrelevant.
It turns out that the definition \ydef\ will ultimately prove
to be
very convenient because $y_l(-\tau) = y_l(\tau)^*$.}
$\theta_l$ is defined by
\eqn\thetadef{
e^{2 i \theta_l} = (-1)^{l + 1} {\Gamma(i \mu) \Gamma(l + 1 - i \mu) \over
\Gamma(-i \mu) \Gamma(l + 1 + i \mu)}.
}
The quantity $\mu \equiv \sqrt{m^2 - 1}$ must be real in order
for $\phi_{lm}$ and $\phi_{lm}^*$ to be interpreted in the usual way
as positive
and negative frequency modes, respectively.\foot{The analysis still
goes through for $0<m^2<1$, although the case $m^2 =0$
is quite subtle \TolleyGG\ and will not be considered here.}
Note that as in \BMS\ we
find it convenient to
use a nonstandard basis of spherical harmonics.
We define
\eqn\ourylm{
Y_{lm} = \sqrt{i \over 2} S_{lm} + (-1)^l \sqrt{- {i \over 2}} S_{lm}^*
}
in terms of the usual spherical harmonics $S_{lm}$, their utility for
our purpose being that they satisfy
\eqn\ylmcool{
Y^*_{lm}(\Omega) = (-1)^l Y_{lm}(\Omega) = Y_{lm}(\Omega_{\rm A}).
}
The modes \globalmodes\ are normalized with respect to the Klein-Gordon
inner product
\eqn\ipglobal{
\langle \phi_{lm}, \phi_{l'm'}\rangle = i (\cosh \tau)^2 \int d^2\Omega\ 
(\phi_{lm}^* \overleftrightarrow{\p_\tau} \phi_{l'm'}) = \delta_{ll'}
\delta_{mm'}
}
by virtue of the fact that
\eqn\useful{
i (\cosh\tau)^2 (
y_l^* \overleftrightarrow{\partial_\tau}
y_{l}) = 1
}
for all $l$.

The phase $e^{2 i \theta_l}$ will play an important role below,
so we record some of its properties.  We define
\eqn\deltadef{
\Delta_\pm(\Omega,\Omega') = -{1 \over \mu \sinh \pi \mu}
\sum_{lm} Y_{lm}(\Omega) Y_{lm}(\Omega')
e^{\mp 2 i \theta_l},
}
which is just the two point function for a conformal field
of dimension $h_\pm \equiv 1 \pm i \mu$ on the sphere \BMS.
It is clear that they satisfy
\eqn\trivial{
(\mu \sinh \pi \mu)^2
\int d^2 \Omega''\ \Delta_-(\Omega,\Omega'') \Delta_+(\Omega'',
\Omega') = \delta^2(\Omega,\Omega').
}

Next we discuss the bulk-boundary
propagators\foot{Although de Sitter space itself has no boundary,
$\scri^+$ and $\scri^-$ are the boundaries of the conformal compactification
of de Sitter space.},
which are used to
construct bulk solutions of the Klein-Gordon equation corresponding
to wavepackets coming in from $\scri^-$ or going out to $\scri^+$.
We define $K^{\pm}$ by
\eqn\kpm{
K^\pm(\Omega';\tau,\Omega) = \sum_{lm}  Y_{lm}(\Omega')
K^\pm_{lm}(\tau,\Omega), \qquad
K_{lm}^\pm(\tau,\Omega) = e^{\pm i \theta_l}
\sqrt{\mu \over 2} Y_{lm}^*(\Omega) y_l(\tau).
}
These are related by
\eqn\kpmrel{
K^\pm(\Omega'; \tau,\Omega) = -\mu \sinh \pi \mu
\int d^2 \Omega''\ \Delta_\mp(\Omega',
\Omega'') K^\mp(\Omega_{\rm A}''; \tau,\Omega)
}
and satisfy
\eqn\aaa{
K^\pm(\Omega';\tau,\Omega) = K^\pm(\Omega;\tau,\Omega') =
K^\pm(\Omega'_{\rm A}; \tau,\Omega_{\rm A}).
}
They are solutions of the wave equation with the boundary
conditions
\eqn\aaa{
\lim_{\tau \to \pm\infty} K^\pm(\Omega';\tau,\Omega)
= e^{(\mp 1 - i \mu) \tau} \delta^2(\Omega,\Omega') + {\cal{O}}(e^{\mp 3
\tau}),
}
i.e. they are positive frequency solutions which approach
delta functions on $\scri^\pm$.
For this reason we will frequently use the notation $K^{\rm in}
\equiv K^-$ and $K^{\rm out} \equiv K^+$.
Given any smooth function $f(\Omega)$ on the sphere, we can construct
solutions of the bulk Klein-Gordon equation by the prescription
\eqn\kpresc{
\phi_f^{\rm in,out}(\tau,\Omega) = \int d^2\Omega'\ f(\Omega')
K^{\rm in,out}(\Omega'; \tau,\Omega).
}
The solution $\phi_f^{\rm in}$ represents a wavepacket with envelope $f$
coming in from $\scri^-$, while $\phi_f^{\rm out}$ represents a wavepacket
with envelope $f$ going out to $\scri^+$.

\subsec{Planar Coordinates}

A basis of positive frequency
solutions of the free Klein-Gordon equation is given by
\eqn\planarmodes{
\phi_{\vp}(t,\vx) = e^{i \vp \cdot \vx} u(p,t), \qquad
u(p,t) = {t J_-(p t) \over \sqrt{8 \pi \sinh \pi \mu}}.
}
We use the notation $J_\pm(z) \equiv
J_{\pm i \mu}(z)$, where $J_\nu(z)$ is
the Bessel function.
Thus $\phi(t) \sim t^{1 - i \mu}$ near $t = 0$.
The modes \planarmodes\ are normalized according to
\eqn\planarnorm{
\langle \phi_{\vp}, \phi_{\vp'} \rangle
= {i \over t} \int d^2x\ (\phi_{\vp}^* \overleftrightarrow{\partial_t}
\phi_{\vp'}) = \delta^2(\vp - \vp').
}
The bulk-boundary propagator 
to $\scri^-$
is
\eqn\bbpropplanar{
K(\vy; t, \vx) = {1 \over 2 \pi} \int d^2p\ e^{i \vp \cdot \vy}
\widetilde{K}(\vp; t,\vx), \qquad \widetilde{K}(\vp; t,\vx) =  e^{-i
\vp \cdot \vx}  z(p) u(p,t),
}
where
\eqn\zdef{
z(p) ={1 \over 2 \pi}
 (p/2)^{i \mu}\Gamma(1 - i \mu) \sqrt{8 \pi \sinh \pi \mu}.
}
This factor will play as important a role as the phase \thetadef\ in
global coordinates.
Performing the Fourier transform \bbpropplanar\ gives
\eqn\aaa{
K(\vy; t,\vx) = -{i \mu\ \over \pi}\theta(t - |\vx - \vy|)
\left( {t \over t^2 - |\vx - \vy|^2}\right)^{1 + i \mu}
}
The solution to the free Klein-Gordon equation corresponding
to an incoming wavepacket with profile $f(\vy)$ from $\scri^-$ is then just
\eqn\phifplanar{
\phi_f(t,\vx) = \int d^2 y\ f(\vy) K(\vy;t,\vx).
}

\newsec{Vacuum States in de Sitter Space}

It was shown
by Breitenlohner and Freedman \BreitenlohnerJF\  that
in AdS${}_{d+1}$, a scalar field whose mass lies in the range
$-({d \over 2})^2 < m^2 < -({d \over 2}
)^2 + 1$ admits two inequivalent quantizations.
Such scalars were later found to play an interesting role
in the AdS/CFT correspondence \refs{\KlebanovTB, \MincesZY}.
Since this phenomenon is related to the fact that for this range of $m^2$,
both mode solutions of the Klein-Gordon equation are normalizable
\BalasubramanianSN, one might expect a similarly interesting story
in the dS/CFT correspondence, where both modes are normalizable for any value
of $m^2$.

In fact, as discussed in
\refs{\ChernikovZM, \TagirovVV,
\AllenUX, \MottolaAR} and reviewed in \BMS, the story is even
richer for de Sitter: there
is a one complex parameter family of $SO(d,1)$ invariant vacuum states
for a massive scalar field in $d$-dimensional de Sitter spacetime (of
which a one real parameter subset are CPT invariant \BMS).
Two vacuum states play special roles:
the Euclidean vacuum $|{\rm E}\rangle$
is the one obtained by analytically continuing from the sphere to
de Sitter spacetime, and the 
$|{\rm in}\rangle$ vacuum is the one with no particles on
$\scri^-$\foot{It was shown in \BMS\ that in odd dimensional de Sitter
spacetime, $|{\rm in}\rangle = |{\rm out}\rangle$, the state with no
particles on $\scri^+$.}.

It is occasionally said that the hypothesized dS/CFT correspondence
is `just' an analytic continuation of AdS/CFT.
However, it is easy to see that analytic continuation from AdS does not give
any vacuum state for a scalar field in dS.  Consider the AdS commutator
function $[\Phi(x),\Phi(y)]$.  It vanishes outside the lightcone in AdS,
but analytic continuation to dS involves 
interchanging the roles of $t$ and $r$\foot{AdS
and dS can both be obtained from Euclidean AdS with metric
${1 \over x_0^2} (d x_0^2 + \cdots + d x_d^2)$, the only difference
being whether one takes $(x_0,x_1) \to (i t,r)$ or $(r, i t)$.},
which turns light cones on their sides.  So simply analytically continuing
the two-point function for a scalar field would give a commutator function
which vanishes inside the light cone, but not outside.
This would violate causal propagation.

\subsec{The MA Transform}

The vacuum associated to the global coordinate modes
\globalmodes\foot{By this we mean the vacuum
annihilated by the operators multiplying \globalmodes\ in the
free field expansion of $\Phi$.} is called the
$|{\rm in}\rangle$ vacuum since it corresponds to having
no particles coming in from $\scri^-$.  
Now consider the frequency independent (i.e., diagonal)
Bogolyubov transformation
\eqn\ma{
\widetilde{\phi}_{lm} = 
{1 \over \sqrt{1 - e^{\gamma + \gamma^*}}}
(\phi_{lm} - e^\gamma \phi_{lm}^*).
}
Following \BMS, we call \ma\ an MA transform for Mottola-Allen
\refs{\MottolaAR,\AllenUX}.
The modes \ma\ define a de Sitter invariant
vacuum state $|\gamma\rangle$ for
any complex $\gamma$ with ${\rm Re}(\gamma)<0$\foot{If
${\rm Re}(\gamma)>0$ then we can simply exchange $\phi$ and $\phi^*$.}.
The two-point function $\langle \gamma| \Phi(x)
\Phi(y)| \gamma\rangle$ has the usual singularity whenever $x$
and $y$ are null separated, but in general it has
an additional singularity whenever
$x$ is null separated from $y_{\rm A}$.
Since this second pole is always separated from $x$ by
a horizon, there is no obvious reason to discard these vacuum states.
The value $\gamma = - \pi \mu$ is the Euclidean vacuum, and $\gamma =
-\infty$ is the $|{\rm in}\rangle$ vacuum.
The two-point function in the Euclidean vacuum has no antipodal
singularity.

\subsec{Two-point Functions in the $|\gamma\rangle$ Vacua}

In this section we record the two-point function $\langle \gamma| \Phi(x)
\Phi(y)|\gamma \rangle$ in the $|\gamma\rangle$ vacuum for later use.
It is straightforward to derive a general identity for the
Wightman two-point function \BMS
\eqn\wightman{
G^{\rm W}_\gamma(x,y) = {1 \over 1 - e^{\gamma + \gamma^*}}
\left[ G_{\rm in}^{\rm W}(x,y)
- e^{\gamma^*}
G_{\rm in}^{\rm W}(x,y_{\rm A}) + e^{\gamma + \gamma^*}
G_{\rm in}^{\rm W}(y,x)  - e^\gamma G_{\rm in}^{\rm W}(x_{\rm A},y)\right]
}
in terms of the Wightman function in the $|{\rm in}\rangle$ vacuum.
It will be convenient to
write an explicit formula,
expressed in terms of the de Sitter invariant
quantity $P$ associated to two points \SpradlinPW.  In global coordinates,
\eqn\globalP{
P(\tau,\Omega;\tau',\Omega') =  \cosh \tau \cosh \tau'
\cos \Theta(\Omega,\Omega')
- \sinh \tau \sinh \tau',
}
where $\Theta$ is the angle between $\Omega$ and $\Omega'$ on $S^2$,
while in planar coordinates
\eqn\planarP{
P(t,\vx;t',\vx') = 1 + {(t - t')^2 - |\vx - \vx'|^2 \over 2 t t'}.
}
It is useful to keep in mind the following properties:
$P(x,y)$ is greater than $1$, equal to $1$, or less than $1$
respectively if $x$ and $y$ are timelike, null, or spacelike separated.
Furthermore $P(x,y) = - P(x,y_{\rm A})$ so that $P(x,y)$ is greater than
$-1$, equal to $-1$, or less than $-1$  respectively if $x$ and $y_{\rm A}$
are spacelike, null, or timelike separated.

In terms of the de Sitter invariant quantity $P$ we can write
the commutator function
\eqn\commutator{
i G^{\rm C}(x,y) \equiv [\Phi(x),\Phi(y)] = 
-{i\over 2 \pi }\,{\rm sign}(x^0 - y^0){\cos[ \mu \cosh^{-1}(P)]
\over \sinh[\cosh^{-1}(P)]}, \qquad P>1.
}
Here ${\rm sign}(x^0 - y^0)$ is $+1$ if $x$ is in the future light cone
of $y$, and $-1$  if $x$ is in the past light cone of $y$.
Of course $G^{\rm C}$ vanishes for spacelike separation, $P<1$.

The commutator function is a $c$-number which is independent of the
state $|\gamma\rangle$ \fulling, so we can summarize the $\gamma$
dependence of the two-point function by looking at the Hadamard
function, which turns out to be 
\eqn\hadamard{
\eqalign{
G^{\rm H}_{\gamma}(x,y) &\equiv
\langle \gamma | \{ \Phi(x),\Phi(y)\} |\gamma\rangle
\cr
&=
- {1 \over \pi} {1 \over 1 - e^{\gamma + \gamma^*}} {{\rm Im} \exp[
\gamma- i \mu \cosh^{-1}(-P)] \over
\sinh [\cosh^{-1}(-P)]},\qquad\qquad\qquad\qquad~ P<-1,\cr
&= 
{\cosh[\mu \cos^{-1}(P)] - \cosh[{\rm Re}(\gamma)]
\cosh[\mu(\pi - \cos^{-1}(P))]
 \over 2 \pi \sinh \pi \mu \sinh[{\rm Re}(\gamma)]},
~~~ -1<P<1,
\cr
&= {\coth[{\rm Re}(\gamma)] \over 2 \pi}{ \sin [ \mu \cosh^{-1}(P)]
\over  \sinh[\cosh^{-1}(P)]},
~~~~~\qquad\qquad\qquad\qquad\qquad\qquad~~ P>1.
}}
Note that 
only the $P<-1$ part is sensitive to the imaginary part
of $\gamma$.
The time-ordered correlation function
\eqn\gfglobal{
\eqalign{
G^{\rm F}(x,y)_\gamma&\equiv
\langle \gamma | T\, \Phi(x) \Phi(y)|\gamma\rangle
\cr
&= \sum_{lm} \theta(\tau  - \tau') \widetilde{\phi}_{lm}(x)
\widetilde{\phi}_{lm}^*(y) + \theta(\tau' - \tau) \widetilde{\phi}_{lm}(y)
\widetilde{\phi}_{lm}^*(x)
}}
will play
the central role beginning in section 5, and the representation
\gfglobal\ will prove more useful than the expression obtained after
the sum is performed.

\newsec{The MA$'$ Transform in Planar Coordinates}

Previous analyses of these vacua have focused on global coordinates, where
the calculations are simpler but the physical meaning of the Euclidean vacuum
is obscure.
In planar coordinates there is a new natural vacuum state: the one defined
by having no particles on the horizon at $t = \infty$.
However, the planar coordinate system only makes a subgroup of the
full de Sitter isometry group manifest.  In particular, the location
of the horizon
at $t=\infty$
is not invariant under de Sitter transformations, so
one might have expected that the boundary condition 
of having no particles on the horizon could not give rise to a de Sitter
invariant vacuum state.  In this section we prove the slightly
surprising result that the one parameter family of vacua do
appear naturally in planar coordinates, and that the state
with no particles on the horizon is just the Euclidean vacuum.

The MA transformation
\ma\ cannot be done on the planar modes \planarmodes\ since
the resulting vacuum state would break translation invariance along
$\vx$ --- the cross terms between $\phi$ and $\phi^*$ would give rise
to terms in the two point function depending on $|\vx + \vy|$ instead
of $|\vx - \vy|$.
To remedy this problem,  consider a modified transformation (which
we call MA$'$) of the
form
\eqn\maprime{
\widetilde{\phi}_{\vp}(t,\vx) =
{1 \over \sqrt{1 - e^{\gamma
+ \gamma^*}}} ( \phi_{\vp}(t,\vx) - e^\gamma \phi^*_{-\vp}(t,\vx))\equiv
e^{i \vp \cdot \vx} \tilde{u}(p,t),
}
where
\eqn\utildedef{
\tilde{u}(p,t) = {1 \over \sqrt{1 - e^{\gamma + \gamma^*}}}(u(p,t)
- e^\gamma u^*(p,t)).
}
Although this does seem
to preserve translation invariance, it is far from obvious that
\maprime\ leads to a de Sitter invariant vacuum state for any
complex $\gamma$ (we again stick to ${\rm Re}(\gamma)<0$), but we 
will now see that this is indeed the case.

Let us start with $\gamma = -\infty$, so that $\widetilde{\phi} = \phi$.
Since the modes \planarmodes\ are purely positive frequency on $\scri^-$,
we expect that if they define any de Sitter invariant vacuum at all,
it should be the $|{\rm in}\rangle$ vacuum.
To check this we calculate the Wightman two-point function
\eqn\wone{
\eqalign{
G^{\rm W}(t,\vx;t',\vx') &=
 \int d^2p\ \phi_{\vec{p}}(t,\vec{x})
\phi_{\vec{p}}^*(t',\vec{x}')
\cr
&= {tt' \over
8 \pi \sinh \pi \mu} \int d^2p\ 
e^{i \vec{p} \cdot (\vec{x} - \vec{x}')}
J_{-}(p t) J_{+}(p t')\cr
&= { t t' \over 4 \sinh \pi \mu}
 \int_0^\infty dp\ p J_0(p|\vec{x}-\vec{x}'|)
 J_{-}(p t) J_{+}(p t').
}}
It is straightforward but tedious to massage 
this integral \gr\ to obtain the result
\eqn\bcdh{\eqalign{
G^{\rm W} &=
0, \qquad\qquad\qquad\qquad\qquad
\qquad\qquad\qquad\qquad\qquad\qquad ~~P < -1,\cr
&={1 \over 4 \pi \sinh \pi \mu} {
\cosh[\mu(\pi - \cos^{-1}(P))]
\over \sin[\cos^{-1}(P)]},\qquad\qquad\qquad -1<P<1,\cr
&= - {i \over 4 \pi}\,{\rm sign}(t-t')
{\exp[-i\mu\,{\rm sign}(t - t') \cosh^{-1}(P)]
\over \sinh[\cosh^{-1}(P)]}, ~~\qquad P > 1,
}}
with $P$ given by \planarP.  From
\bcdh\ we find an expression for
$G^{\rm C} = 2\,{\rm Im}(G^{\rm W})$ which agrees with \commutator,
and we find that the Hadamard function
\eqn\inhadamard{
\eqalign{
G^{\rm H} = 2\,{\rm Re}(G^{\rm W}) &= 0,
~~~~~~~~\qquad\qquad\qquad\qquad\qquad\qquad\qquad\qquad P<-1,\cr
&= {1 \over 2 \pi \sinh \pi \mu}
{\cosh[\mu(\pi - \cos^{-1}(P))]
\over \sin[\cos^{-1}(P)]}, \qquad -1<P<1,\cr
&=
- {1 \over 2 \pi} { \sin[ \mu \cosh^{-1}(P)]
\over \sinh \left[ \cosh^{-1}(P)\right]},
~~~~\qquad\qquad\qquad\qquad P>1
}}
agrees with \hadamard\ for $\gamma = -\infty$.
This concludes the proof that the planar modes
\planarmodes\ define the $|{\rm in}\rangle$ vacuum.

Now consider arbitrary $\gamma$ in \maprime.
Since the commutator function is unaffected by this
Bogolyubov transformation, we need only to calculate
\eqn\abc{
G^{\rm H}_{\gamma}
= 2\,{\rm Re} \int d^2p\ \widetilde{\phi}_\vp(t,\vx)
\widetilde{\phi}_{\vp}^*(t',\vx').
}
Using \maprime, we can write
\abc\ as
\eqn\hadplanar{
G^{\rm H}_{\gamma} =
{1 \over 1 - e^{\gamma + \gamma^*}}\left[
 (1 + e^{\gamma + \gamma^*})
G^{\rm H}_{{\rm in}}
- 4\,{\rm Re}(e^{\gamma^*} I)\right],
}
with $G^{\rm H}_{{\rm in}}$ given by \inhadamard\ and
\eqn\idef{
I = {t t' \over 8 \pi \sinh \pi \mu} \int_0^\infty d^2p\  e^{i \vp \cdot
(\vx - \vx')} J_-(p t) J_-(p t').
}
It is again straightforward to check that \hadplanar\ is equal to
\hadamard.

Finally we address the significance of the Euclidean vacuum from the
point of view of the MA$'$ transform.  Plugging $\gamma = -\pi \mu$
into \maprime, we find that the modes which give the Euclidean vacuum are
\eqn\precis{
\phi^{\rm E}_{\vp}(t,\vx) = {t e^{i \vp \cdot \vx} 
\over \sqrt{8 \pi \sinh \pi \mu}} {1 \over \sqrt{1 - e^{-2 \pi \mu}}}
\left[ J_-(p t) - e^{- \pi \mu} J_+(p t)\right].
}
Using the asymptotic expansion of the Bessel function we find that
\eqn\asympt{
\phi^{\rm E}_{\vp}(t,\vx) \sim 
{1 \over 2 \pi} \sqrt{i t \over 2 p}
e^{i \vp \cdot \vx - i p t}
}
near $t = \infty$.
We see that precisely that linear combination \precis\ which gives the
Euclidean vacuum is the one which is purely positive frequency near
the horizon.  
This shows that the Euclidean vacuum is  natural 
for cosmological purposes, when one might want to put boundary conditions
on the scalar field on the past horizon of the region ${\cal{O}}^+$.

Before concluding our discussion of planar coordinates, we record
here the momentum space Feynman propagator $G^{\rm F}$,
which will be used in the calculations below:
\eqn\gfplanar{
G^{\rm F}_{\gamma}(x,x') = \int d^2 p\ e^{i \vp \cdot (\vx - \vx')}
\widetilde{G}^{\rm F}_\gamma(t,t',p),
}
where we define the function
\eqn\gfplanar{
\widetilde{G}^{\rm F}_{\gamma}(t,t',p) = 
  \theta(t-t') \tilde{u}(p,t) \tilde{u}^*(p,t')
+ \theta(t'-t) \tilde{u}^*(p,t) \tilde{u}(p,t').
}

\newsec{dS/CFT in Global Coordinates}

After the existence of an AdS/CFT correspondence was suggested by
Maldacena \MaldacenaRE, a precise prescription for calculating
CFT correlation functions in terms of AdS data was soon developed
\refs{\GubserBC, \WittenQJ}.  In AdS/CFT, the gravity partition function,
viewed as a functional of boundary data, serves as the generating
functional of CFT correlators,
\eqn\adscft{
Z[\phi_0] = \left<\exp\ {i \int_{\p {\cal{M}}} \CO \phi_0}\right>.
}

Recently the dS/CFT correspondence has been proposed by
Strominger \StromingerPN\ and a recipe for
calculating two-point CFT correlation functions
has been suggested, but a precise
dictionary between bulk and boundary correlation functions has not been given.
There are at least two related
reasons why adopting the AdS prescription
is problematic.

As mentioned above, all of the modes in de Sitter space are
normalizable, but in Lorentzian AdS, normalizable and non-normalizable
modes play substantially different roles \BalasubramanianSN.
The former encode the states of the theory while the
latter correspond to boundary conditions for fields
and do not fluctuate.
The AdS boundary conditions ensure,
for example, that the on-shell action for a scalar
field, which is a total derivative $S = \int dz\ \p_z(z^{n-1} \phi \p_z \phi)$,
has only a contribution from the boundary $z = 0$ and not from the horizon
$z = \infty$.
But in dS, there is no boundary condition (other than the trivial one
$\phi(t,\bx) \equiv 0$)
one could impose on $\phi$ at $\scri^-$ in order to eliminate
the contribution to the on-shell action from the horizon at $t=\infty$.
This highlights the necessity of having two independent CFT operators
for every bulk field $\phi$, as opposed to the 1-1 correspondence familiar
from AdS.  (This argument applies in planar coordinates 
--- of course in global coordinates one also expects two CFT operators,
simply because there are two boundaries $\scri^\pm$.)
The fact that two CFT operators are associated to each bulk field
has indeed been discussed in \StromingerPN.

In evaluating the on-shell
action in AdS space there are divergences which are easily regulated
by prescribing boundary conditions not at $z=0$ but at $z=\epsilon$.
A well-defined result is obtained after subtracting the power-law
and logarithmic divergences as $\epsilon \to 0$.
In dS a similarly regulated on-shell action for a scalar
field is not infinite but does not converge --- it 
has terms on $\scri^-$ which behave like $e^{i/\epsilon}$
as $\epsilon \to 0$.  In planar coordinates there are as mentioned in
the previous paragraph also non-zero terms coming from the horizon
which behave as $e^{i T}$ as $T \to \infty$.
We have been unable to find a regularization
scheme which enables one to extract sensible results.

We proceed by recalling
an alternative interpretation of the same 
AdS/CFT correlation functions
which 
was developed in \refs{\BalasubramanianDE,
\PolchinskiRY, \SusskindVK,
\BalasubramanianRI,
\ArefevaMI}
and nicely proven in \GiddingsQU.
Giddings showed that the CFT correlators calculate the $S$-matrix
for scattering particles from the boundary of AdS into the bulk and back.
In AdS/CFT, the boundary CFT has
a timelike direction, and the positive and
negative frequency components of a CFT operator
${\cal{O}}_\phi$
create
and annihilate quanta of the associated bulk field $\phi$.

We propose to adopt a suitable generalization of the construction of
\GiddingsQU\
to define CFT correlators for de Sitter space. 
Since the boundary conformal field theory is Euclidean,  instead
of having positive and negative frequency components of the operator
we need  the two operators ${\cal{O}}_+$, $\cal{O}_-$.
In fact, in global coordinates it makes sense to think about four
different operators ${\cal{O}}^{\rm in, out}_+$ and
$\cal{O}^{\rm in,out}_-$.
We interpret ${\cal{O}}^{\rm in}_+$ and ${\cal{O}}^{\rm in}_-$
as coupling respectively to positive
and negative
frequency quanta of the bulk field $\phi$ on $\scri^-$.
On $\scri^+$ the pairing is reversed:  ${\cal{O}}^{\rm out}_+$
couples to negative frequency quanta, and ${\cal{O}}^{\rm out}_-$
couples to positive frequency quanta.  This convention ensures
that operators ${\cal{O}}_\pm$ have conformal weight $h_\pm = 1 \pm i \mu$
regardless of whether they are in or out operators.
Only two of the four operators are independent, and we will discuss below
how to relate the out operators to the in operators perturbatively.

Concretely, our proposal is to define dS/CFT correlation
functions in global coordinates by
the prescription
\eqn\corrdef{
\eqalign{
&\langle 
\prod_{i=1}^m \CO^{\rm out}_+(\Omega_i) \prod_{j=1}^n
\CO^{\rm in}_+(\Omega'_j)
\rangle
= \lim_{{\tau_i \to +\infty} \atop {\tau_j' \to -\infty}}
\int \Big[
\prod_{i=1}^m (\cosh \tau_i)^2 d^2 \omega_i \  K^{\rm out *}(\Omega_i;x_i)
i \overleftrightarrow{\partial_{\tau_i}}
\Big]\cr
&\qquad
\qquad
\qquad
\times
G^{\rm F}(x_1,\ldots,x_m,x_1',\ldots,x_n')
\Big[\prod_{j=1}^n  (\cosh \tau_j')^2  d^2 \omega_j'\ 
i  \overleftrightarrow{\partial_{\tau_j'}}
 K^{\rm in}(\Omega_j';x_j')
\Big],
}}
where $G^{\rm F}$ is the bulk time-ordered Feynman correlation function
and we use the notation $x = (\tau,\omega)$ (additional
details of \corrdef\ will be clarified below).

The formula \corrdef\ only defines the two operators
${\cal{O}}^{\rm in,out}_+$, but it is straightforward to generalize
the prescription to include the other two operators.  Schematically,
for
every insertion of ${\cal{O}}^{\rm in}_-$ we include
\eqn\oinminus{
\langle \cdots {\cal{O}}^{\rm in}_-(\Omega) \cdots \rangle = 
\lim_{\tau' \to -\infty} \int (\cosh \tau')^2 d^2\Omega'\   K^{{\rm in}*}
(\Omega; x') i \overleftrightarrow{\partial_{\tau'}} G^{\rm F}(\cdots,
x', \cdots),
}
while an insertion of ${\cal{O}}^{\rm out}_-$ involves
\eqn\ooutminus{
\langle \cdots {\cal{O}}^{\rm out}_-(\Omega) \cdots \rangle = 
\lim_{\tau' \to +\infty} \int (\cosh \tau')^2 d^2\Omega'\ 
G^{\rm F}(\cdots, 
x', \cdots) i \overleftrightarrow{\partial_{\tau'}} K^{\rm out}(\Omega; x').
}
The ordering of the operators inside these correlation functions is
irrelevant, except for possible contact terms, which
can be computed explicitly (as we will show below).

The motivation for the proposal \corrdef\ comes from studying
$S$-matrix elements in \dst.
In the next subsection we will derive an LSZ-like
formula for the $S$-matrix and show that it can be written in
terms of the correlation functions
\corrdef\ as
\eqn\dscftdef{
S[\{f_i\}; \{g_j\}] =
\int \Big[\prod_{i=1}^m {d^2\Omega_i
\over \sqrt{Z}}
f_i^*(\Omega_{i})
\Big]
\Big[
\prod_{j=1}^n {d^2 \Omega'_j
\over \sqrt{Z}}
g_j(\Omega'_j)\Big]
\langle 
\prod_{i=1}^m \CO^{\rm out}_+(\Omega_i) \prod_{j=1}^n \CO^{\rm in}_+(\Omega'_j)
\rangle,
}
where
$f_i$ and $g_j$ are  smooth functions on the sphere,
and the left-hand side is the $S$-matrix element for $n$ incoming
wavepackets with envelopes $f_i$ and $m$ outgoing wavepackets with
envelopes $g_j$.
The factor $Z$ is a wavefunction renormalization which 
one could calculate
perturbatively.

\subsec{Motivation: the $S$-matrix}

Following standard arguments \Bjorken, we consider the interaction
of some wavepackets which are widely separated in the
far past and in the far future, so that the full
interacting field $\Phi(x)$ asymptotes to free fields,
\eqn\asympt{
\lim_{\tau \to -\infty} \Phi(x) = \sqrt{Z} \Phi^{\rm in}(x), ~~~~~
\lim_{\tau \to +\infty} \Phi(x) = \sqrt{Z} \Phi^{\rm out}(x).
}
Here we allow for a wavefunction renormalization $Z$, and the canonically
normalized free fields $\Phi^{\rm in,out}$ are expanded in terms
of operators $a^{\rm in,out}$ as
\eqn\free{
\Phi^{\rm in,out} = \sum_{lm} \phi_{lm} a^{\rm in,out}_{lm} + \phi_{lm}^*
a^{{\rm in, out}\dagger}_{lm}, \qquad
a^{\rm in,out}_{lm}|0\rangle = 0.
}
Note that we are taking the in and out vacua to be the
same, as is appropriate for \dst\ \BMS.
The role of the choice of vacuum will be discussed below.
The condition \asympt\ holds weakly (i.e., it is not an operator
identity but is valid inside matrix elements).  The
operators $a^{\rm in,out}$ are recovered from the free fields
$\Phi^{\rm in,out}$ by the standard formula
\eqn\aaa{
a^{\rm in,out}_{lm}
= 
\int (\cosh \tau)^2
d^2\Omega\ \phi^*_{lm}(\tau,\Omega) i \overleftrightarrow{\partial_\tau}
\Phi^{\rm in,out}(\tau,\Omega).
}
Since $\Phi^{\rm in}$ and $\Phi^{\rm out}$ both satisfy the free
wave equation,
these operators are independent of $\tau$.

Given a smooth function $g(\Omega)$ on $\scri^-$, we can use the
bulk-boundary propagator $K^{\rm in}$ of subsection 2.1 to construct
a solution $\phi_g^{\rm in}$ of the wave equation which represents an
incoming wavepacket with envelope $g$, as in equation 
\kpresc.
This wavepacket corresponds to the state $\alpha_g^{{\rm in}\dagger}|0
\rangle$, where
\eqn\ainoutdef{
\alpha_f^{\rm in,out} \equiv  \int (\cosh \tau)^2 d^2 \Omega \ 
\phi_f^{{\rm in,out}*}
(\tau,\Omega) i \overleftrightarrow{\partial_\tau} \Phi^{\rm in,out}(\tau,
\Omega).
}

Similarly, an outgoing wavepacket with envelope $f$ at $\scri^+$ is
constructed using $K^{\rm out}$, and corresponds to the state
$\langle 0 | \alpha_f^{\rm out}$.
The $S$-matrix element for $n$ incoming wavepackets $\{g_j\}$ and
$m$ outgoing wavepackets $\{f_i\}$ is defined by
\eqn\smatrixdef{
S[\{f_i\};\{g_j\}] = \langle 0| \prod_{i=1}^m \alpha_{f_i}^{\rm out}
\prod_{j=1}^n \alpha_{g_j}^{{\rm in}\dagger}|0\rangle.
}

Now using the definitions
\ainoutdef\ \smatrixdef\ and the asymptotic condition
\asympt, it is straightforward to derive a formula for the $S$-matrix
\eqn\smatrix{
\eqalign{
 S[\{f_i\}; \{g_j\}] &=
\lim_{{\tau_i \to +\infty} \atop {\tau_j' \to -\infty}}
\int \Big[ \prod_{i=1}^m (\cosh \tau_i)^2 {d^2 \Omega_i\over
\sqrt{Z}}  \ \phi^{\rm out *}_{f_i}(x_i)
i   \overleftrightarrow{\partial_{\tau_i}}
\Big]
\cr
&\qquad
\times
\langle 0|T\ \prod_{i=1}^m
\Phi(x_i) \prod_{j=1}^n \Phi(x_j')
|0\rangle
 \Big[ \prod_{j=1}^n (\cosh \tau_j')^2 {d^2 \Omega_j' \over \sqrt{Z}}\ 
i \overleftrightarrow{\partial_{\tau_j'}}
\phi^{\rm in}_{g_j}(x_j')
\Big].
}}
Note that the derivative operators do not hit the factors of
$(\cosh \tau)^2$ in the measure, as in \ainoutdef.
Also, the time-ordering symbol
inside the bulk correlation function can be interpreted as defining
the order in which the $\tau$ coordinates are taken to infinity.
One should evaluate the quantity at a fixed $\tau_1 > \cdots
\tau_m > \tau_1' > \cdots >\tau_n'$ and then take the limits
preserving that ordering.  In particular, one need not worry
about delta function contributions coming from when the
$\tau$ derivatives hit the time-ordering symbol.
Note that the vacuum $|0\rangle$ in \smatrix\ can be any of the
vacuum states discussed in the previous sections\foot{In fact, there
is no reason to necessarily take the incoming and outgoing vacua
to be the same, although we will not pursue this possibility here.}.
We will see by explicit calculation how the $S$-matrix elements, and
hence the CFT correlators, depend on this choice of vacuum.

So far we have only used the operators ${\cal{O}}_+^{\rm in,out}$,
but it is clear how to introduce the other two.  
In the $S$-matrix \smatrixdef\ we can also include
operators like $\alpha^{{\rm in}}$ and $\alpha^{{\rm out} \dagger}$, which
lead straightforwardly to the prescriptions \oinminus, \ooutminus\ for
insertions of ${\cal{O}}_-^{\rm in}$ and ${\cal{O}}_-^{\rm out}$
respectively.  The possibility of including these operators may seem
unfamiliar since normally one can use only $a^{{\rm in} \dagger}$ and
not $a^{\rm in}$ for constructing initial states since the latter
annihilates $|0\rangle$.    But we can include them here since we
will be interested in the $\gamma$-dependence of the CFT correlation
functions, and $a^{\rm in}$ only annihilates $|\gamma\rangle$ for
$\gamma = -\infty$, the $|{\rm in}\rangle$ vacuum.

\subsec{Relation Between in and out Operators}

Using Green's theorem and the fact that $K^{\rm out}(\Omega; x)$ satisfies
the equation of motion $(\nabla^2_x - m^2 ) K^{\rm out} = 0$ immediately
gives the formula
\eqn\outinrel{\eqalign{
&-i \int \sqrt{-g}\,dx'\,K^{\rm out}(\Omega; x') (\nabla_{x'}^2 - m^2)
G^{\rm F}(\cdots, x', \cdots)
\cr
&\qquad\qquad= (\lim_{\tau' \to +\infty}
- \lim_{\tau'\to -\infty})
\int d^2 \Omega' \ i (\cosh \tau')^2 
G^{\rm F}(\cdots,x',\cdots)
\overleftrightarrow{\partial_{\tau'}}
K^{\rm out}(\Omega; x').
}}
Now the first term on the right-hand side looks like an insertion
of ${\cal{O}}^{\rm out}_-$, as in \ooutminus, while the second term
on the right-hand side can be made to look like an insertion of
${\cal{O}}^{\rm in}_+$ by recalling the relation \kpmrel.  This leads
to the identity
\eqn\relation{
\eqalign{
\langle \cdots {\cal{O}}^{\rm out}_-(\Omega) \cdots \rangle
&= - \mu \sinh \pi \mu \int d\Omega'\ \Delta_-(\Omega, \Omega'_{\rm A})
\langle \cdots {\cal{O}}^{\rm in}_+(\Omega') \cdots \rangle
\cr
&\qquad\qquad
- i \int \sqrt{-g} dx'\ K^{\rm out}(\Omega; x') (\nabla_{x'}^2 - m^2)
G^{\rm F}(\cdots, x', \cdots).
}}
Of course a similar formula relates ${\cal{O}}_-^{\rm in}$ and
${\cal{O}}_+^{\rm out}$.
For  two-point functions in 
the free theory, it is not hard to see that the second line of
\relation\ vanishes, so that one obtains the weak operator identities
\eqn\opident{
{\cal{O}}^{\rm out}_\pm(\Omega) =
- \mu \sinh \pi \mu \int d\Omega'\ \Delta_\pm(\Omega, \Omega'_{\rm A})
{\cal{O}}^{\rm in}_\mp(\Omega').
}
This relation receives perturbative corrections which can in principle
be derived from the identity \relation.

\subsec{CFT Two-point Functions}

We now show that the proposal \corrdef\
reproduces
the two-point functions of \BMS\ in an arbitrary vacuum.
This calculation is trivial in momentum space, so we start by Fourier
transforming \corrdef\ to obtain
\eqn\aaa{\eqalign{
\langle \prod_{i=1}^m {\cal{O}}^{\rm in}_{- l_i m_i}
\prod_{j=1}^n {\cal{O}}^{\rm in}_{+ l_j' m_j'} \rangle &=
\lim_{{\tau_i \to -\infty} \atop {\tau_j' \to -\infty}}
\Big[
\prod_{i=1}^n
k^*_{l_i}(\tau_i)
i (\cosh \tau_i)^2 \overleftrightarrow{\partial_{\tau_i}}\Big]
\cr
&\times
{\tilde{G}}^{\rm F}_{ \{ l_i m_i\}, \{l_j',m_j'\}}( \{\tau_i\}, \{\tau_j'\})
\Big[
\prod_{j=1}^m
i (\cosh \tau_j')^2 \overleftrightarrow{\partial_{\tau_j'}}
k_{l_j'}(\tau_j')
\Big],
}}
where we have defined
\eqn\kdef{
k_l(\tau) \equiv
e^{-i \theta_l} \sqrt{\mu \over 2} y_l(\tau).
}
This is essentially just $K^{\rm in}$,
but with the spherical harmonics stripped off
already.

Since $k_l(\tau)$ is just $y_l(\tau)$ up to a factor, it is trivial
to use the momentum space representation \gfglobal\ and the orthogonality
of the modes \ipglobal\ to obtain
\eqn\aaa{
\langle
\gamma|{\cal{O}}^{\rm in}_{-lm} {\cal{O}}^{\rm in}_{-l'm'}
|\gamma\rangle
= - {e^{\gamma^*} \over 1 - e^{\gamma + \gamma^*}}
e^{2 i \theta_l} {\mu \over 2},\qquad {\rm etc.}
}
Fourier transforming back to position space gives
\eqn\gtwopoints{\eqalign{
\langle\gamma| \CO^{\rm in}_-(\Omega) \CO^{\rm in}_-(\Omega')|\gamma\rangle &= {\mu^2 \over 2}
\sinh \pi \mu
{e^{\gamma^*}
\over 1- e^{\gamma + \gamma^*}}
\Delta_-(\Omega, \Omega'),\cr
\langle\gamma| \CO^{\rm in}_+(\Omega) \CO^{\rm in}_+(\Omega')|\gamma\rangle &= {\mu^2 \over 2}
\sinh \pi \mu
{e^{\gamma}
\over 1- e^{\gamma + \gamma^*}}
\Delta_+(\Omega, \Omega'),\cr
\langle\gamma| {\cal{O}}^{\rm in}_-(\Omega) {\cal{O}}^{\rm in}_+(\Omega')|\gamma\rangle
&= {\mu \over 2} {1 \over 1 - e^{\gamma + \gamma^*}} \delta^2(\Omega,
\Omega'),
\cr
\langle\gamma| {\cal{O}}^{\rm in}_+(\Omega) {\cal{O}}^{\rm in}_-(\Omega')|\gamma\rangle
&= {\mu \over 2} {e^{\gamma + \gamma^*} \over 1 - e^{\gamma + \gamma^*}} \delta^2(\Omega,
\Omega'),
}}
in agreement with the results of \BMS\ (after translating from our
$\gamma$ conventions to their $\alpha$ conventions).

\subsec{CFT Three-point Function}

Of course as far as the two-point functions \gtwopoints\ are concerned,
one could eliminate the $\gamma$ dependence by rescaling the operators
${\cal{O}}^{\rm in,out}_\pm$.
In this section we outline the calculation of a CFT three-point function
in the presence of a $\phi^3$ interaction in the bulk, and prove
that an invariant ratio of correlation functions
depends nontrivially on $\gamma$.
This provides evidence that these $|\gamma \rangle$
vacua are marginal deformations of the CFT, as opposed to simply
field rescalings.
The calculation appears more difficult than the
corresponding calculation in
AdS/CFT\foot{
In planar coordinates, the technical difficulty arises because the
three-point function involves integrals like
$\int d^2 y\ K(\vx_1;\vy) K(\vx_2;\vy) K(\vx_3;\vy)$,
but the $\theta$-function in
the bulk boundary propagator $K$ \bbpropplanar\ makes this integral
difficult to manipulate.  In particular, the clever AdS tricks of \fred\ do
not seem to work.},
but fortunately we will
be able to exploit the simple behavior of the global coordinate modes
under the antipodal map to extract the essential features of the
result.
The invariant ratio we will calculate is
\eqn\threeone{
R(\gamma)\equiv
{\langle \gamma | {\cal{O}}_+^{\rm in} (\Omega_1)
{\cal{O}}_+^{\rm in}(\Omega_2) {\cal{O}}_+^{\rm in}(\Omega_3)|\gamma
\rangle^2 \over
\langle \gamma | {\cal{O}}_+^{\rm in} (\Omega_1)
{\cal{O}}_+^{\rm in}(\Omega_2)|\gamma\rangle
\langle \gamma | {\cal{O}}_+^{\rm in} (\Omega_1)
{\cal{O}}_+^{\rm in}(\Omega_3)|\gamma \rangle
\langle \gamma | {\cal{O}}_+^{\rm in} (\Omega_2)
{\cal{O}}_+^{\rm in}(\Omega_3)|\gamma \rangle
}.
}

Since our calculation will not be able to determine the overall
($\gamma$-independent) constant in $R$, we omit overall constants
throughout this calculation.  The prescription \corrdef\ amounts to
extracting the coefficient of $e^{h_+(\tau_1 + \tau_2 + \tau_3)}$
as all three points approach $\scri^-$.  That is,
\eqn\aaa{
\lim_{\tau_i \to -\infty}
G_\gamma^{\rm F}(x_1,x_2,x_3) \sim e^{h_+(\tau_1 + \tau_2 + \tau_3)}
\langle \gamma| {\cal{O}}^{\rm in}_+(\Omega_1) {\cal{O}}^{\rm in}_+(\Omega_2) 
{\cal{O}}^{\rm in}_+(\Omega_3) |\gamma\rangle + \cdots.
}
Diffeomorphism invariance of $G_\gamma^{\rm F}(x_1,x_2,x_3)$ ensures
that the CFT three-point function read off in this manner will be
conformally invariant.

At tree level in perturbation theory we have
\eqn\threetwo{
G_\gamma^{\rm F}(x_1,x_2,x_3) = \int \sqrt{-g}\,dx\ G^{\rm F}_\gamma(x,x_1)
G^{\rm F}_\gamma(x,x_2)
G^{\rm F}_\gamma(x,x_3).
}
Since we are interested in the limit
$\tau_i \to -\infty$, it is safe to replace the time-ordered two-point
functions in \threetwo\ by the Wightman function (it is not hard
to check carefully that the difference of the integrals goes to zero
in the limit we are interested in).  Then we use the identity
\wightman, but note that the second and third terms in \wightman\ behave
as $e^{h_- \tau'}$ near $\scri^-$ and therefore do not contribute to
\threeone.
Keeping only the terms which behave like $e^{h_+(\tau_1 + \tau_2 + \tau_3)}$
gives
\eqn\threefour{
(1 - e^{\gamma + \gamma^*})^{-3} \int \sqrt{-g}\,dx \prod_{i=1}^3
\left[ G^{\rm W}(x,x_i) - e^\gamma G^{\rm W}(x_{\rm A},x_i)\right].
}
The eight terms in \threefour\ can
easily be combined by noting that in every integral with two or three
$x_{\rm A}$'s we can make the change of variables $x \to x_{\rm A}$ to
end up with only one $x_{\rm A}$ or none.  Therefore \threefour\ is
equal to
\eqn\threefive{
(1 - e^{\gamma + \gamma^*})^{-3} \left[ (1 - e^{3 \gamma})
G - e^\gamma(1 - e^\gamma) (G_1+G_2+G_3)\right],
}
where
\eqn\threesix{
G\equiv \int \sqrt{-g}\,dx\ G^{\rm W}(x,x_1)
G^{\rm W}(x,x_2)
G^{\rm W}(x,x_3)
}
and
\eqn\threeseven{
G_1 \equiv \int \sqrt{-g}\,dx\ G^{\rm W}(x_{\rm A},x_1)
G^{\rm W}(x,x_2)
G^{\rm W}(x,x_3), \qquad {\rm etc}.
}
In fact it is safe to replace $G^{\rm W}$ by the time-ordered
product $G^{\rm F}$ in \threesix\ and \threeseven\ since we are
only interested in the limit $\tau_i \to -\infty$.  In any case,
diffeomorphism invariance of the integrals \threesix\ and
\threeseven\ implies that coefficients of
$e^{h_+(\tau_1 + \tau_2 + \tau_3)}$
must be proportional to the conformally invariant three-point function
$\Delta_{+++}$
for a field of weight $h_+$.
We have not determined the constants of proportionality, but since
there is no $\gamma$-dependence in the remaining integrals
\threesix\ and \threeseven,
the $\gamma$-dependence of  \threefour\ must be of the form
\eqn\aaa{
\langle \gamma| {\cal{O}}_+^{\rm in} (\Omega_1)
{\cal{O}}_+^{\rm in}(\Omega_2) {\cal{O}}_+^{\rm in}(\Omega_3)|\gamma
\rangle \sim  (1 - e^{\gamma + \gamma^*})^{-3}
\left[ x (1 - e^{3 \gamma})  - y e^\gamma (1 - e^\gamma) \right]
\Delta_{+++},
}
where $x$ and $y$ are undetermined non-zero constants.
We conclude that
the invariant ratio \threeone\ is
\eqn\aaa{
R(\gamma) \sim e^{-3 \gamma} (1-e^{\gamma+ \gamma^*})^{-3}
\left[x (1 - e^{3 \gamma}) -  y e^\gamma (1 - e^\gamma)\right]^2.
}
Although we have not determined $x$ or $y$, it is clear that no choice
renders $R(\gamma)$ independent of $\gamma$.  Therefore we
conclude that the $\gamma$ dependence of the CFT correlation functions
cannot be absorbed into a rescaling of the operators ${\cal{O}}$.

\newsec{dS/CFT in Planar Coordinates}

In planar coordinates,
we propose to define dS/CFT correlation functions
by the rule
\eqn\planardef{
\eqalign{
\langle \prod_{i=1}^m {\cal{O}}_-(\vx_i) \prod_{j=1}^n
{\cal{O}}_+(\vy_j)\rangle
&= \lim_{t_i, t_j' \to 0} \int \Big[
\prod_{i=1}^m {d^2 x_i \over t_i} \ K^*(\vx_i; x_i) i
\overleftrightarrow{\partial_{t_i}}
\Big]
\cr
&~~~
\times
G^{\rm F}(x'_1,\ldots,x'_m;y'_1,\ldots,y'_n) \Big[
\prod_{j=1}^n {d^2 y_j' \over t_j'}\ i \overleftrightarrow{\partial_{t_j'}}
K(\vy_j'; y_j')
\Big],
}}
with the notation $x = (t,\vx)$ and $y = (t',\vy)$.
Again the ordering of the operators is irrelevant except for contact
terms, which can be computed by ordering the $x_i'$ and $y_j'$ in
parallel with the corresponding $\vx_i$ and $\vy_j$.
In the next subsection we motivate this definition by analyzing
the $S$-vector \WittenKN.

\subsec{Motivation: The $S$-vector}

In planar coordinates covering ${\cal{O}}^-$ it does not make sense to
speak of asymptotic out states since the horizon is located at a
finite affine distance from any point in the bulk of ${\cal{O}}^-$.
(In our formalism, this problem manifests
itself through the lack of a `bulk-horizon' propagator which one
could use to propagate wavepackets from the horizon.)
Therefore it has been proposed \WittenKN\ (see also \HellermanYI)
that the natural meta-observable is not the $S$-matrix 
but an $S$-vector, where a unique state $\langle U|$
is generated by some unknown mechanism
on the horizon, and the only calculable quantities
are $\langle U| a\rangle$, for states $|a\rangle$ on the boundary
(which we take to be $\scri^-$).
This is the point of view we will adopt, although
for simplicity we will only consider the case when $\langle U|$
is one of the de Sitter invariant vacuum states $|\gamma\rangle$.

We define the $S$-vector
\eqn\svector{
S[\{f_i\}; \{g_j\}] = \langle 0 | \prod_{i=1}^m
\alpha_{f_i}^{\rm in} \prod_{j=1}^n \alpha_{g_j}^{{\rm in}
\dagger} |0\rangle,
}
where
\eqn\aaa{
\alpha_f^{\rm in} = {i \over t} \int d^2 x\ \phi_f(t,\vx)
\overleftrightarrow{\p_t} \Phi^{\rm in}(t,\vx),
}
with $\phi_f$ defined in 
\phifplanar.  We have kept the superscript ``in'' in 
these formulas to compare with the previous section, but since here there is
no ``out'', they will henceforth be dropped.
Note that the operator $\alpha_{f}$ in \svector\ annihilates
a wavepacket with envelope $f$ at $\scri^-$.  This makes sense
because a general de Sitter invariant vacuum state $|0\rangle$
which we might choose to use in \svector\ actually contains
an infinite number of particles on $\scri^-$.
We form the initial 
state by adding the wavepackets $g_j$ and deleting the
wavepackets $f_i$ from this state.

Repeating the LSZ analysis of the previous section, it is easy
to write the $S$-vector in the form
\eqn\aaa{
S[\{f_i\};\{g_j\}] = \int \Big[ \prod_{i=1}^m
{d^2 x_i \over \sqrt{Z}} f_i^*(\vx_i)\Big]
\Big[\prod_{j=1}^n {d^2 y_j \over \sqrt{Z}} g_j(\vy_j)\Big]
\langle \prod_{i=1}^m {\cal{O}}_-(\vx_i) \prod_{j=1}^n
{\cal{O}}_+(\vy_j)\rangle,
}
with the CFT correlator on the right hand side given
by \planardef.

\subsec{CFT Two-point Functions}

In momentum space, the CFT correlation function \planardef\ is
simply
\eqn\plmom{
\eqalign{
\langle \prod_{i=1}^m {\cal{O}}_-(\vp_i)
\prod_{j=1}^n {\cal{O}}_+(\vq_j)\rangle
&= \lim_{t_i,t_j' \to 0} \Big[ \prod_{i=1}^m k^*(p_i,t_i) {i \over t_i}
\overleftrightarrow{\partial_{t_i}}
\Big] 
\cr
&\qquad\qquad\times
G^{\rm F}(\{ t_i,\vp_i \}, \{t_j',\vq_j\})
\Big[\prod_{j=1}^n {i \over t_j'} \overleftrightarrow{
\p_{t_j'}} k(q_j,t_j')\Big],
}}
with
$k(p,t) \equiv 2 \pi z(p) u(p,t)$, recalling \bbpropplanar\ and \zdef.

Using \gfplanar\ and the orthogonality \planarnorm\ of
the modes, we find
immediately
\eqn\aaa{
\eqalign{
\langle \gamma|{\cal{O}}_-(\vx) {\cal{O}}_-(\vy) |\gamma\rangle
&=-
{e^{\gamma^*} \over 1 - e^{\gamma + \gamma^*}}
\int d^2p\ e^{i \vp \cdot (\vx - \vy)} (2 \pi z^*(p))^2
\cr
&= -16 i (\pi \mu)^2 {e^{\gamma^*} \over 1 - e^{\gamma + \gamma^*}}
{1 \over |\vx - \vy|^{2 h_-}}
}}
and
\eqn\aaa{
\eqalign{
\langle\gamma| {\cal{O}}_-(\vx) {\cal{O}}_+(\vy)
|\gamma
\rangle &= {1 \over 1 - e^{\gamma + \gamma^*}}
\int d^2p \ e^{i \vp \cdot (\vx - \vy)} |2 \pi z(p)|^2\cr
&=
{1 \over 1 - e^{\gamma + \gamma^*}}
2 \mu (2 \pi)^4 \delta(\vx - \vy).
}}

\newsec{Summary and Discussion}

The purpose of this paper has primarily been to define a procedure
for calculating CFT correlation functions from bulk $n$-point
functions in \dst.  Although our proposal is modeled on
a similar procedure from AdS/CFT, we have highlighted some of
the important differences between dS and AdS which make naive
extrapolation of AdS results impossible.
These differences include the fact that in dS one inevitably
has two CFT operators for every bulk field $\phi$ (since there
is no natural boundary condition one could impose to eliminate the
second operator), as well as the fact that a scalar field in de Sitter
space has a whole family of different vacuum states, none of which
is the one obtained from AdS by analytic continuation.

We have also shown
that these de Sitter invariant vacuum states arise naturally in
coordinates covering only half of de Sitter space, where the Euclidean
vacuum plays the special role of having no particles on the horizon.
Finally, we have sketched the calculation of a CFT three-point function
and shown that an invariant ratio \threeone\ of correlation functions 
depends nontrivially on the choice of vacuum $\gamma$.
This shows that the $\gamma$ dependence of the CFT correlation
functions cannot be eliminated by rescaling
the operators. However, it leaves open the intriguing
possibility that the correlation functions
may be related by a $\gamma$--dependent {\it nonlocal}
field redefinition of ${\cal {O}}_\pm^{{\rm in},{\rm out}}.$
This is easily seen to be true for the two-point functions
\gtwopoints,
and it would be interesting to see whether this is a general feature.

Our $S$-matrix and $S$-vector proposals answer the question of what
these CFT correlation functions
of \refs{\WittenKN, \StromingerPN, \BMS} are.
Unfortunately, we have not answered the interesting and pressing
question of how to interpret these quantities, which have been
called `meta-observables' \WittenKN\ since no single observer in
de Sitter space can
access more than a single point on $\scri^+$.
Also, in this formulation of the dS/CFT, the CFT lives on a Cauchy
surface at infinite distance,
rather than a boundary.  It might be more satisfactory,
from a holographic point of view, to have a formulation in which the CFT
lives on the horizon \SachsQB.

\vskip .7cm

\centerline{\bf Acknowledgements}

\vskip .2cm

We have benefited greatly from discussions with R. Bousso, I. Klebanov,
A. Maloney, I. Savonije, K. Skenderis, E. Verlinde and H. Verlinde.
We are especially grateful to A. Strominger for
many useful discussions and to the authors of \BMS\
for early drafts.
M.S. is supported by DOE grant DE-FG02-91ER40671, and
A.V. is supported by DE-FG02-91ER40655.

\listrefs

\bye